\documentclass{eas}   
\usepackage{graphicx}
\usepackage{natbib}
\usepackage{wrapfig}

\def\Msun {\hbox{$M_\odot$}}

\def\etal {et al.}

\def\HII  {\hbox{H{\sc ii}}}

\def\UCHII  {\hbox{UCH{\sc ii}}}

\def\apj{{ApJ} }
\def\apjl{{ApJL} }

\def\aap{{A\&A} }
\def\mnras{{MNRAS} }
\def\araa{{ARA\&A} }

\def\apjs {{ApJS}}
\def\apss {{Astrophys. and Space Sc.}}

\def\nat {{Nature}}

\begin{document}   
   
\title{High-Mass Star Formation} 
\author{Peter Schilke}\address{I. Physikalisches Institut der Universit\"at zu K\"oln, Z\"ulpicher Str. 77, 50937 K\"oln, Germany, schilke@ph1.uni-koeln.de}
\begin{abstract}   
A review on current theories and observations of high-mass star formation is given.  Particularly the influence of magnetic fields and feedback mechanisms, and of varying initial conditions on theories are discussed.  The, in my biased view, most important observations to put strong constraints on models of high-mass star formation are presented, in particular bearing on the existence and properties of high-mass starless cores, the role of filaments in the mass transport to high-mass cores, and the properties of disks around high-mass stars.   
\end{abstract}
\maketitle
\section{Introduction}

\begin{figure} 
\centering
\includegraphics[bb=0 220 600 600, width=0.7\textwidth]{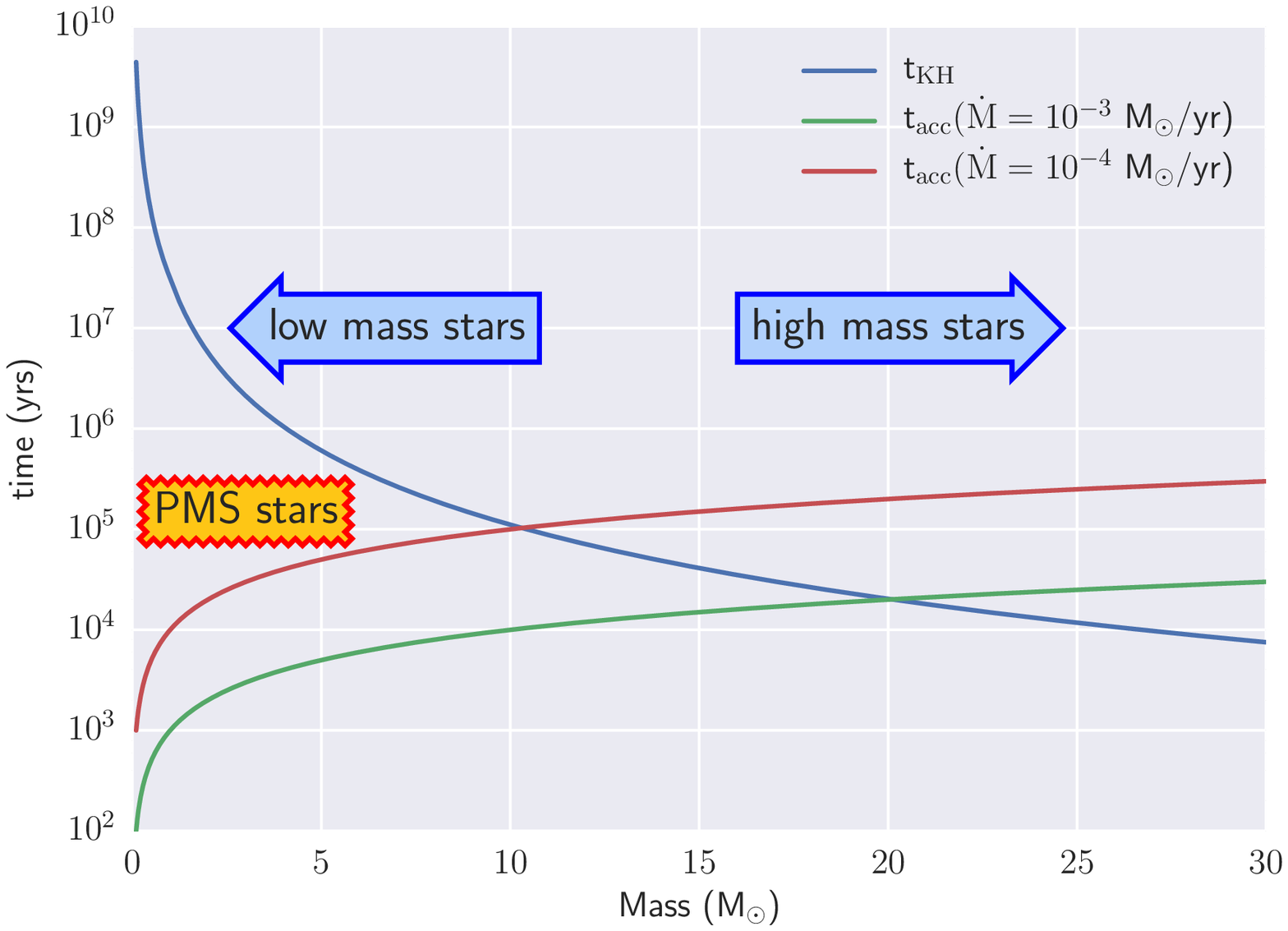}

\caption{Kelvin-Helmholtz time scale vs. accretion time scale of high mass stars vs. low mass stars, for varying accretion rates.  Only low mass stars have a pre-main sequence phase, where they have stopped accreting, but get their energy from gravitational contraction. 
} \label{schilke:fig1}
\end{figure}

High-mass stars dominate the energy input in galaxies, through mechanical (outflows, winds, supernova blast waves) and radiation (UV radiation creating  \HII\ regions) input.  Particularly through supernova explosions, they can shape whole galaxies \citep{Bolatto2013}. They also enrich the ISM with heavy elements, which in turn modifies the star formation process. Yet, their formation process will differ from low-mass stars in significant ways: while the Kelvin-Helmholtz timescale of low-mass stars is significantly longer than the time required to assemble them, for any reasonable accretion rate it is shorter for high-mass stars (Fig.~\ref{schilke:fig1}).  This has as a consequence that high-mass stars above a certain mass (again depending on the accretion rate) will have to continue accreting after reaching the main sequence.  
This means that the feedback of the stars in the later stages of accretion in form of radiation is considerable.  \citet{WolfireCassinelli1987} could show that spherical accretion with normal grains could make the creation of stars of more than 10 \Msun\ impossible, because the radiation pressure on dust would halt accretion.  As stars more massive than 10 \Msun\ do exist, this led to various attempts to circumvent this barrier.

\section{Theories of High-Mass Star Formation}

\citet{WolfireCassinelli1987} speculated about different dust properties. A later insight was to abandon the assumption of spherical accretion.  Any angular momentum of the infalling gas will lead to the creation of a disk, which funnels accretion in the equatorial plane with much higher rates per area.  At the same time, most of the radiation escapes in the polar regions, and therefore cannot interact with the infalling dust particles.  This was investigated by \citet{YorkeSonnhalter2002}.

\subsection{Turbulent core vs. competitive accretion}



But even with spherical accretion, this barrier can be overcome if the accretion rates are high enough (see Fig.~5 from \citealt{WolfireCassinelli1987}). If the accretion rate depends only on thermal support, \citet{Shu77} find typical values of 10$^{-5}$ \Msun/yr, which is not sufficient for forming high-mass stars.  \citet{McKeeTan2003} therefore considered turbulent support, which allows much higher accretion rates.  They call their model ``turbulent core'', but it is also known as ``monolithic collapse''.  It is a very influential paper, and frequently used. Part of its appeal is that it is analytical, so it can make many easily accessible predictions on derived values (e.g.\ mass vs. time, or density profiles as function of time).  It does however make some quite strong assumption on initial conditions. Apart from the spherical approximation, it starts with a strongly peaked density distribution ($n\propto r^{-1.5}$). It is therefore fully applicable if and only if such strongly peaked cores exist. One characteristic of this model is that the final stellar mass is pre-assembled in the collapsing core (even if it may form binary or multiple stars), i.e.\ it assumes that the clump mass is isolated from the rest of the cloud and directly maps into the initial stellar mass distribution.

An alternative approach was taken by \citet{Bonnell01}.  They considered star formation in a cluster.  All stellar embryos are created with equal mass, and then gather mass through Bondi-Hoyle accretion.  Their accretion rate therefore is determined by their location in the cluster potential -- near a gravitational well, where more mass is available, a star can accrete more, in the outskirts of the cloud, less.   Gravitational interactions between stars can kick them out of rich feeding zones, which terminates their accretion.  Since the stars have to compete for their resources, this model is called ``competitive accretion''. It uses a uniform gas density as initial condition, and the original model was isothermal.  In this model, the final star can, depending on its trajectory, draw from the vast mass reservoir of the whole cloud.  Here, there is no connection between the mass of its birth core and the final stellar mass.

Turbulent core and competitive accretion were almost orthogonal approaches to the problem of high-mass star formation, and to a detached observer, they shared the same type of pitfalls: both used very specific and somewhat arbitrary, if very different, initial conditions (although they were physically motivated), and both lacked important physics, mostly related to feedback.  It therefore was to be expected that both would capture some aspects of the real star formation process, but that none of them would be the final word.  Yet, for quite a while, there was a very heated debate, and in many observing proposals the two models were described as mutually incompatible, and only one of them being right, which would be decided by the observations proposed. In the following, I discuss the progress made since these original models, by including more physics, and it can be seen that the descendants of the models are not as far apart as their ancestors were.  It is to be expected that they will merge into a unified star formation theory as time goes by.

\subsection{Magnetic Fields}
One problem not mentioned so far is the fact that isothermal clouds would fragment to quite small masses, if only thermally supported.  The turbulent core model tackles this problem by assuming additional turbulent support.  Competitive accretion circumvents it by assuming that indeed all stellar embryos are small, but then get the mass in a different way. None of the two models used magnetic fields, for computational simplicity.  While it is widely assumed, based on observations, that molecular clouds are supercritical \citep{Crutcher2012}, i.e.\ that the magnetic field is not strong enough to stop cloud collapse, it may still affect it.  And indeed, \citet{Hennebelle2011} find that by including magnetic fields, the fragmentation is reduced by a factor of 2, i.e.\ the cloud fragments into fewer, but higher mass fragments.  Another important effect of magnetic fields is the magnetic braking that removes angular momentum from the densest part of the cores, and therefore also affects star formation.

\subsection{Feedback Processes}

Feedback processes from already existing stars inject energy and momentum into the remaining cloud in different ways:  mechanical (outflows, supernovae), and  radiative (thermal, ionizing) feedback.  Outflows inject kinetic energy and momentum into the cloud (see e.g.\ \citealt{Wang10}), and can therefore sustain turbulence opposing infall, as well as disrupt mass flows through filaments.  It is unclear however, how much of the energy and momentum of such outflows stays in the cloud, and how much is lost once the outflow pierces the boundary of the cloud and escapes. But the net effect is that outflow feedback reduces the infall rates, and therefore hinders the formation of high-mass stars. Supernova feedback operates on GMC scales, and will terminate all star formation in its immediate vicinity by removing the gas mass reservoir.

Thermal feedback is a very important feedback mechanism.  Stars emit radiation, thereby heat up their environment, which raises the Jeans mass ($M_J \propto T^{3/2} n^{-1/2}$), and results in fewer, but higher-mass fragments and hence stars. This was investigated in detail e.g.\ by  \citet{Krumholz2007} and \citet{Bate2009}.  Here, just the thermal feedback was investigated.  However, high-mass stars emit quite large numbers of EUV photons, which are capable of ionizing hydrogen.  One would naively think that this is devastating to further growth.  However, if one considers real geometries including angular momentum, one finds that accretion proceeds mostly through disks, which means that all the mass flow is channeled into a relatively small area, which intercepts only a small fraction of the incident UV flux.  Most of the radiation escapes through the polar regions and does not significantly interact with infalling gas.  This is known as flashlight effect \citep{Krumholz2009, Peters2010}.  Also relevant in this context is that accretion can proceed through ionized gas \citep{Keto2007}.

\subsection{Initial Conditions}
As discussed above, not only did the early models miss some important physics, but the initial conditions were very different: while the turbulent core model used a highly peaked density structure, the density structure of the molecular cloud of the competitive accretion model was flat. It seemed plausible that these different intial conditions had influence on the results, and this was systematically investigated by \citet{Girichidis2012}. They ran the same models with the same physics, but different initial conditions, and investigated the resulting cluster structure.  Indeed it was found that peaked distributions resulted in no sub-clustering (just like turbulent core models), while flat density distributions resulted in strong sub-clustering (just like competitive accretion).  

Therefore, one of the strongest preconditions for getting realistic models of high-mass star formation is to get the initial conditions right.  This is not trivial, because the initial conditions of star formation are the results of cloud formation models, which in turn depend on their own initial conditions, determined by galaxy formation and evolution models, etc. Since these models all deal with vastly different scales (from kpc in galaxy models to AU for star formation), it is numerically not possible to calculate this all self-consistently. Instead, one would calculate models on one hierarchical scale, and use this as input for the next smaller scale.  There are a couple of simulations in the literature that lend themselves to such zoom-in studies: the SILCC simulation that calculates a large section of a galactic disk including supernova feedback, but without a galactic potential (i.e. no spiral arms; \citealt{Walch2015}); other simulations do use a spiral potential, but limited physics, particularly no feedback \citep{Dobbs2013, Dobbs2015, Smith2014}. 

\section{Observations of High-Mass Star Formation}
An alternative, and the ultimate way, to determine the initial conditions, is to use observations.  This poses technical problems though: apart from the ubiquitous one that one observes just a 2-d projection in space and a 1-d projection in velocity, while one needs to reconstruct a 3+3d phase space, high-mass star formation is rare and short-lived.  Thus, there are only a limited number of instances at any given time in a galaxy, and they are, on average, far away.  Moreover, since high-mass star formation happens deeply embedded, only observations at FIR and longer wavelengths can penetrate the cores.  Observations hence require high-resolution instruments in the mm/submm wavelength range.  Only recently has, with ALMA, such an instrument come online, and we can certainly expect many new results in the future.

\subsection{High-Mass Starless Cores}

One of the preconditions for the turbulent core model is the existence of high-mass prestellar cores, i.e.\  objects that have pre-assembled masses sufficient to form high-mass stars in ($> 40 M_\odot$ for the lower end of high-mass stars).  Are there such objects?  \citet{Motte2007} found 129 dense cores in Cygnus X, among them 
40 with masses $> 40 M_\odot$.  Of those, 17 are IR quiet, i.e.\ show no sign of a star having been formed already.  \citet{Bontemps2010} looked at the 5 brightest of these objects and found that all but one are sub-fragmented, that is, they are not forming high-mass stars, but a small cluster of lower mass stars.  The one remaining candidate was shown by \citet{DuarteCabral2013} to have an outflow.  Therefore, star formation is already ongoing, and this core has to be removed from the list of high-mass prestellar mass candidates.  \citet{Tan2014} found a candidate high-mass core of 60\ \Msun, which needs to be supported by strong magnetic field to be stable.  \citet{Kong2015} corroborates a slow collapse of this cloud, using deuterated molecules as a chemical clock. \citet{Cyganowski2014} reports on a very dense core of $\approx 30 \Msun$ showing no sign of star formation.  Based on these observations, it cannot be excluded that genuine high-mass prestellar cores, as required by the turbulent core model, do exist.  They seem to be exceedingly rare, not extremely high-mass (assuming a star-formation efficiency of 30\%, the \citet{Tan2014} core might produce at most one single 18~\Msun\ star, and more likely a multiple system with lower mass stars, or even a small cluster).  Thus it does not appear that monolithic collapse out of pre-assembled centrally peaked high-mass prestellar cores constitutes the dominant mode of high-mass star formation. 

This is a very limited sample though, and by now there are large surveys of the galactic plane.  And indeed, \citet{Csengeri2014} find, in their ATLASGAL sample, that 25\% of their high-mass cores are quiescent.  However, this is based on data taken with a rather large beam, and no test for fragmentation or outflows as early sign of star formation has been performed yet, so this is a strict upper limit.  Systematic studies of these candidates with ALMA at high resolution to establish the fragmentation state and to search for outflows as early signs of star formation are  needed to put the findings on a solid statistical basis.

\subsection{Finding High-Mass Protostars}

\begin{wrapfigure}{L}{0.5\textwidth}
\centering 
\includegraphics[width=0.5\textwidth]{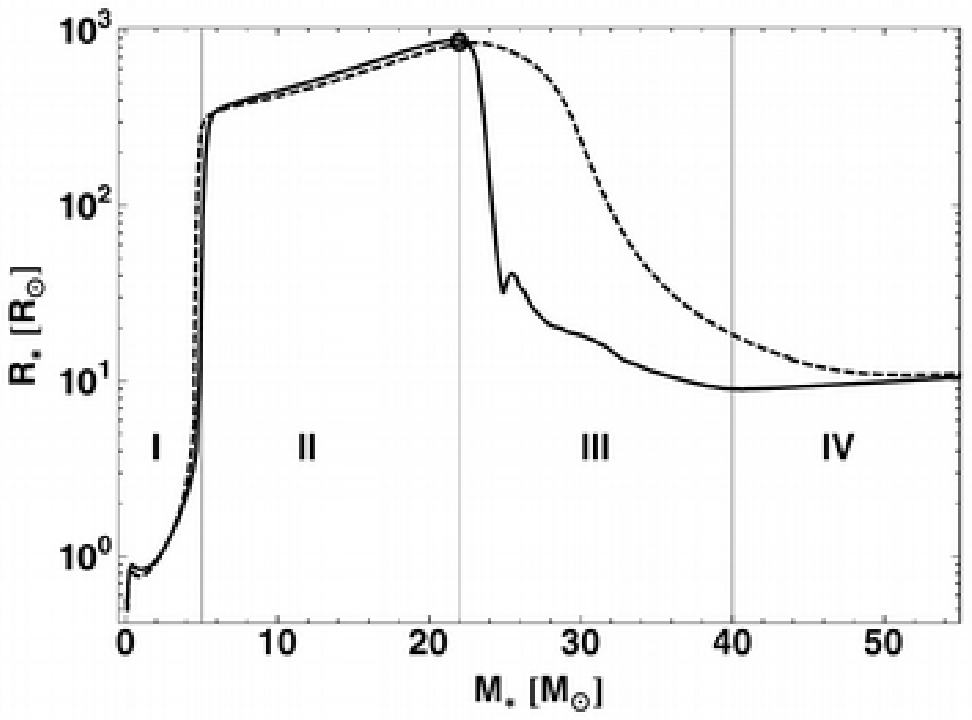}
\includegraphics[width=0.5\textwidth]{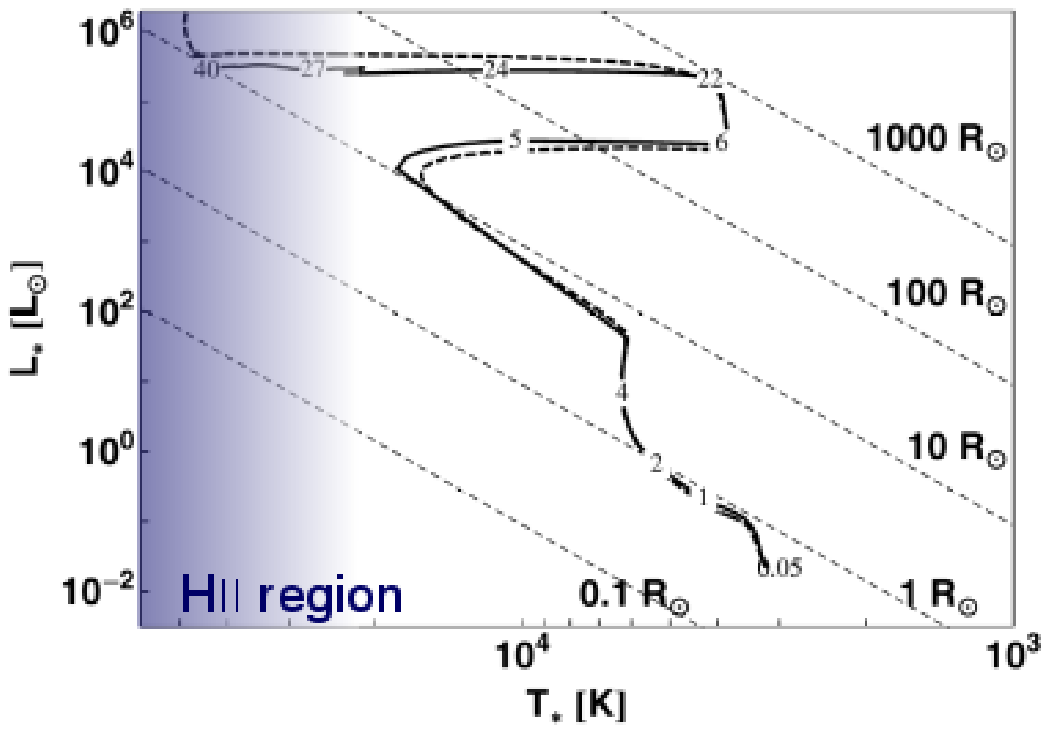}

\caption{Radii of stellar evolution (top) and evolutionary tracks  (bottom) (Fig.~4 and 1, respectively, from \citealt{Kuiper2013}, based on calculations by \citealt{HosokawaOmukai2009} and \citealt{Hosokawa2010} \copyright AAS. Reproduced with permission).} \label{schilke:fig3}
\end{wrapfigure}

Early searches for high-mass protostars used IRAS color-color criteria and then targeted searches for cm continuum radiation from embedded \UCHII\ regions \citep{WoodChurchwell1989}. \UCHII\ regions are readily observable signposts for high-mass stars that already have reached the main sequence, and probably already have accreted most of their mass.  An interesting question therefore is what exactly high-mass protostars will look like, and in particular at which stage one can expect to detect \UCHII. If one assumes that high-mass protostars look just like ZAMS stars, one would expect \UCHII\ to appear when the stars is of spectral type B2, or about 10~\Msun\ \citep{Straizys1981}.   However, the protostars are still accreting mass, and hence do not look like ZAMS stars.  \citet{HosokawaOmukai2009} and \citet{Hosokawa2010} have calculated the stellar structure for high accretion rates ($\approx 10^{-3}$\Msun/yr), and find that for a long time during their development, the protostellar radius is very large, and the protostars are bloated (Fig.~\ref{schilke:fig3}). This means that, although they have a high luminosity, their temperature is too low to emit a sufficient amount of EUV photons, and the \HII\ region does not appear until they have reached about 30~\Msun.  Thus, the observational absence of cm radiation is not necessarily evidence for the absence of a high-mass protostar.

\subsection{Filaments and Mass Flow}
One of the lasting legacies of \emph{Herschel} is the result that filaments are ubiquitous in the interstellar medium, and that star-forming cores are often found at the intersection of filaments \citep[e.g.][]{Schneider2012}. Thus, it was suggested and then shown \citep{Peretto2013, Liu2015} that mass flows along the filaments toward the cores.  Observing this is not trivial, since nothing is known about the orientation of the filaments on the plane of the sky, and velocity gradients cannot be unambiguously connected to infalling or out-streaming gas.  The traditional method of using line profiles to detect mass infall \citep{Evans1999}, which was derived for spherical infall, has been shown by \citet{Smith2013} not to provide unambiguous results, depending on the geometry. 

\subsection{High-Mass Disks} 
As already mentioned, high-mass star formation theories predict accretion disks to sustain accretion in the presence of radiation pressure.  Determining the disk properties observationally thus gives important constraints on models.  There is indirect evidence for disks through the existence of outflows, but since lower mass stars also contribute to outflows, and outflows of clusters tend to overlap, at least in projection \citep{Beuther2003}, they cannot be used to infer disk properties.  The predicted sizes are between 100 and 1000 AU \citep{Kuiper2015}.  There have been reports on disks around B-stars \citep{Cesaroni2005, Cesaroni2014, SanchezMonge2013}, which report sizes of about 2000~AU, and Keplerian rotation.  When the resolution is high enough to resolve the disk, asymmetries are found, attributed to tidal interactions with a central binary \citep{Cesaroni2014}.  

Around O-stars, prior to 2014, only very large rotating structures called toroids were found \citep{Beltran05, Beltran2011}.  These structures are large (10,000 AU), massive (100\Msun) and rapidly contracting, i.e.\ not showing Keplerian rotation.  They are supposed to feed clusters rather than single stars. \citet{Hunter2014, Zapata2015, Johnston2015} present disk candidates around O-stars, with \citet{Johnston2015} showing the best evidence for Keplerian rotation.  The inferred sizes (1000-2000 AU) are at the outer range of expected sizes, but do at present not exclude the possibility that these actually are circumbinary disks.  This would not be surprising, since \citet{Sana2014} find that about 80\% of O-stars have a companion closer than 1000 AU.

\section{Outlook}
The properties most important to study observationally, and that will give the strongest constraints on models (in my biased view) are the following:

\begin{itemize}
\setlength\itemsep{0.0pt}
\item \textbf{Are high-mass cores connected to a larger cloud mass reservoir?}  This can be studied through observations of mass-flow along and onto filaments.
 \item \textbf{What do the initial conditions look like - what is the fragmentation status, what are the density profiles?}  ALMA mapping of high-mass cores and sophisticated modeling (see e.g.\ Schmiedeke et al., this volume) will be necessary.
 \item \textbf{What is the role of magnetic fields in shaping clouds and influencing dynamics?}  This will require dust polarization studies at large and small scales, and Zeeman splitting observations.
 \item \textbf{Are there disks around high-mass stars and what do they look like?}  Here, high-resolution ALMA observations will advance the field within a very short time.
\end{itemize}
A last word of caution:  given the complexity of the star formation process, one does not necessarily expect to find uniform properties.  It is well probable that there are some cores that do resemble the initial conditions for turbulent cores, others may more look like the ones in the competitive accretion scenario.  In some places, magnetic fields may be important, in others, not.  So the ultimate goal would be to first describe the properties discussed above on a statistically sound basis, and then trying to find the causes for the properties being what they are.  
 
\paragraph{Acknowledgements} 
This work has been supported by the Collaborative Research Centre 956, sub-projects A6 and C3, funded by the Deutsche Forschungsgemeinschaft (DFG) and by the German Ministry of Science (BMBF) trough contract 05A11PK3. I thank \'Alvaro S\'anchez-Monge for a critical reading of the manuscript.
\setlength{\bibsep}{0.0pt}

\end{document}